# Synthetic gauge field enabled realization of bulk- and edge-transported states in an aperiodic acoustic structure

Yu-Xin Fang [1,2], Wen-Hao Zhu[1,2], Yuhui Cai[1,2], Xi-Hui Li[1,2], Meng-Qi Zhang[1,2], Jiayao Huang[1,2], Yongyao Li[1,2] and Shi-Qiao Wu[1,2]†

[1]*School of Physics and Optoelectronic Engineering, Foshan University, Foshan 528000, China*
[2]*Guangdong-Hong Kong-Macao Joint Laboratory for Intelligent Micro-Nano Optoelectronic Technology, Foshan University, Foshan 528000, China*

†Correspondence should be addressed to: sqwu@fosu.edu.cn

Topologically protected edge states with immunity against various disorders have been implemented in a variety of topological insulators. In this Letter, we reveal that Landau levels in aperiodic acoustic structures can be achieved under different pseudomagnetic fields (PMFs). The produced zero order Landau modes (ZOLMs) could transmit along the channels at the interior or exterior of the inhomogeneous array, which are separately termed as "bulk-transported states" (BTSs) and "edge-transported states" (ETSs). Distinct from conventional valley edge states, the ZOLMs show intriguing self-collimation feature. If a pseudoelectric field (PEF) is further included, the combination of a PMF and PEF can result in the formation of bulk or edge Landau rainbow, where Landau zero modes are distributed at various positions of the bulk or boundary of the sample at different frequencies. The synthetic-gauge-field-controlled topological states can enable fully control of robust transmission, and using the entire footprint of a topological lattice. Our findings not only profoundly advance the current understanding of topological phase matter but also offer new avenues for constructing topological acoustic devices.

## I. INTRODUCTION

Exploration of controlling the sound flows in new-style acoustic devices contributes to their practical applications for data processing, communication, and computing. Highly integrated transmission for most of the state-of-the-art devices has been an efficient method to improve their utilized efficiency and reduce fabricated costs. However, the boundary transport behaviors of topologically protected acoustic devices not only acquire a bulky lattice even with rare occupation in the bulk but also are more susceptible to external interference, making large-scale integrated applications challenging.

Synthetic gauge fields (SGFs), stemming from nonuniform strain or deformation, provide an alternative approach to investigate the magnetic-like effects even in systems without magnetic field. The Landau modes realized through SGFs are in fact bulk-transported states (BTSs), enabling acoustic components with highly integrated applications possible. Since it has been first reported that in the absence of a magnetic field electrons in a strained graphene sheet can still experience external potential due to SGFs to form effective Landau levels and edge states [1-3]. SGFs and associated intriguing physics have arouse fundamental interest. They have been successively realized in various platforms including electronic graphene [4-6], microwave resonators [7], photonic [8-11] and phononic crystals [12-19]. Furthermore, SGFs and Landau levels also play pivotal roles in understanding topological band theories, giving rise to profound comprehension of the quantum theory and quantum technology [20-22]. In contrast to the genuine magnetic fields, strain-induced pseudomagnetic fields (PMFs) actually come from the spatial shift of the Dirac points in reciprocal space, offering great freedoms to tune the amplitude of PMFs which may suffer huge challenge in real magnetic field. Compared with Landau levels in waveguide arrays [23-25] or multimode cavities [26], the realization of artificial gauge fields in classical wave systems offer new avenue to enhance wave-matter interactions for the flat bands with high degeneracy and high density of states. As a result, it would be highly appealing to provide a new paradigm for on-chip routing and confinement of classical waves.

In artificial acoustic crystals (ACs), Dirac points and Weyl points formed by linear crossing of dispersion relations in momentum space are tightly tied to crystalline symmetry. The singularities are commonly viewed as the topological sources in band theory. A variety of novel phenomena such as topologically protected surface states can be attributed to the emergence of these singular points [27-29]. Their topological properties can be characterized by various topological invariants such as Chern numbers [30-32], spin Chern numbers[33-35], valley Chern numbers [36-38] and bulk polarization [39-41]. The motion of sound near the singular cones can be described by the dynamical equations of relativistic Fermions in quantum field theory. Consequently, such acoustic systems with singular points become good candidates to emulate relativistic particles and observe their peculiar transport behaviors. More recently, people have predicted that systems with singular points can generate flat Landau levels near Dirac frequency while their periodic lattice are deformed [42, 43]. Suitable lattice deformation can remarkably shift the singular points in reciprocal space, acting as the external magnetic field and is termed to as PMFs. PMFs attained in inhomogeneous structure can remarkably break through the limitation of achieving artificial gauge field in periodic structure. Then through engineering the SGFs in artificial systems, it is feasible to attain magnetic-field-like effects for electromagnetic light and sound. This offers an unprecedented approach to control classical waves.

In this Letter, we judiciously design an inhomogeneous structure based on well-known Dirac cone system in a hexagonal lattice consisting of triangular scatterers embedded in air to explore the acoustic realization of SGFs. Rotation angle, distortion and size of the acoustic blocks are three degrees of freedom to construct nonuniform systems. When rotated angle and distortion of the blocks are continuously varied in a manner, PMFs with chiral Landau levels can be successfully achieved near the Dirac frequency. BTSs can be induced by gradually rotating the angle of the blocks while edge-transported states (ETSs) by their gradient deformation. In the two cases, the chiral Landau levels stem from different physical mechanisms. More precisely, rotation angle of the scatterer mainly modulates the spectral gap, and its distortion deviates the Dirac point from their original valley positions. In contrast, the gradient size variation shifts the spectral band up or down overall to induce the formation of PEF. Here we find that the zero order Landau modes (ZOLMs) in the bulk host self-collimating effect to filter the oblique incident wave. The bulk transport has robustness against local disorders and is immune to backscattering. Furthermore, acoustic BTSs can also transmit along different designed channels in the bulk. If multiple SGFs are combined and integrated in the same structure, a novel phenomenon termed Landau rainbow in which Landau modes disperse in various spatial positions at different frequencies can be implemented. With simple structural design without any requirement on external magnetic field or gyromagnetic material, the present proposal for realizing robust BTSs and ETSs is expected to be promising for integrated application in the near future.

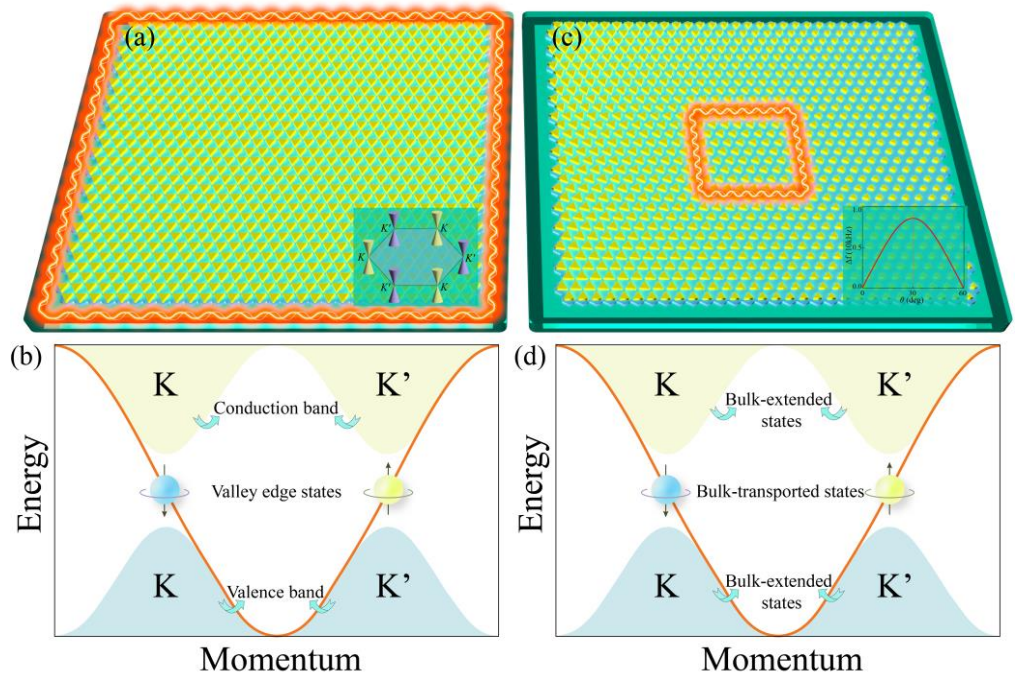

FIG. 1. (a) Schematics of a periodic array with edge states confined at the outer boundaries. It consists of triangular blocks immersed in air. The inset shows the Dirac cones emergent at the corners of the Brillouin zone. (b) Energy diagram as a function of the momentum with gapless valley edge states connecting conduction and valence band within the spectral gap. (c) Illustration of gradually varied building blocks arranged in triangular lattice. The embedded inset exhibits the bandgap evolution of periodically arranged blocks as a function of their rotation angles. Similarly, the aperiodic structure can also produce two spectral bands within the gap as (d) schematically depicts. But the in-gap bands can propagate along the bulk channel.

In many previous works reporting valley Hall effects [44-52], valley edge states can be attained by reducing crystalline symmetry. If the rotation angle of scatterer elements is tuned to deviating from the integer multiple of $60^0$, the degenerate points at the valleys would be lifted to generate valley edge state. And the emergent edge modes typically confine at the outer boundaries of the arrays as marked by wavy lines in Fig. 1(a). At the same time, the band diagram of this periodic lattice would be like the sketch drawing in Fig. 1(b). In great contrast to reported acoustic works with periodicity, here we demonstrate that aperiodic acoustic structure can also produce localized states within the spectral gap, which can be "bulk-transported states" (BTSs) in the bulk or "edge-transported states" (ETSs) at the edges, depending on the direction of SGFs. The highlighted closed loop in the schematic Fig. 1(c) shows the possible transmitted paths of BTS but with in-gap dispersion relations similar to the counterpart of real valley edge states [see Fig. 1(d)].

To better elucidate the forming mechanism of BTSs, we construct hexagonal lattice with triangular blocks immersed in air, wherein the sound velocity and density are respectively $c = 343\ m/s$ and $\rho = 1.23\ kg/m^3$. And lattice constant is chosen as $a = 10\ mm$. We first consider a ribbon-shaped structure with fixed block side length $L = 8.6\ mm$ yet their rotation angles vary from $90^0$ to $30^0$ to form graded morphology as shown in Fig. 2(a), which covers the maximum and minimum bandgap over a rotation periodicity [see the inset in Fig. 1(c)]. The rotation angle of the triangular block is defined as a corner deviating from $x$ axis as the inset of the third spectral map in Fig. 2(a) shows. Such evolution process is sketched by the gradient thickness and colors of a big arrow. Five band diagrams are also plotted for the unit cells indicated by blue tiny arrows. In the supercell, from left to right, spectral bandgap of various unit cells decreases from maximum to zero and return to maximum again, such scenario can enable large bandgap and gapless in-gap LLs. It can be understood by perturbative theory near Dirac point, which characterizes the quasiparticle motion as [53]

$$H = v(k_x\sigma_x \pm k_y\sigma_y) + m\sigma_z \qquad (1)$$

where $v$ represents group velocity, $k_x$ and $k_y$ are the $x$ and $y$ components of wave vectors with respective to $K$ or $K'$ valley. $\sigma_i(i = x, y, z)$ are the Pauli matrices. If rotation angle of rigid blocks takes the values $\theta = n\pi/3$ ($n = 0,1,2 ...$), $C_{3v}$ symmetry is preserved and effective mass $m$ vanishes. Then Eq. (1) degrades into the well-known form for Dirac fermions. However, if $\theta$ deviates from the integer multiple of $\pi/3$, broken parity-inversion symmetry would result in the occurrence of an effective mass $m$. Subsequently, Dirac degeneracy is lifted and a local bandgap emerges near $K$ or $K'$ valleys. The gap size is $\Delta\omega = 2|m|$. The gradient angles in the supercell of Fig. 2(a) induce a gap size $\Delta\omega$ that depends linearly on the inhomogeneous direction, which is equivalent to an effective mass $m$ having a linear relationship with the coordinate $x$ ($m = bx$). As the effective mass $m$ can be regarded as a vector potential, then Eq. (1) equivalently describe a Dirac system under magnetic field $B_y$, which is the so-called SGFs. The effect on Dirac system of SGFs can be revealed by the projected band structure of the inhomogeneous ribbon-shaped supercell with periodic boundary condition in $x$ direction and sound hard boundaries in the remaining regions, which can be numerically calculated by the commercial software COMSOL Multiphysics based on finite element method. As the projected band diagram Fig. 2(b) shows, SGFs produce both higher-order (gray regions) and zero order Landau modes (red curves) in the graded supercell. Though sharing similar trend with the valley edge states inside the spectral gap, ZOLMs are actually bulk states. And the slope and size of the LLs can be illustrated by the formulars as follows [16]

$$\Delta\omega_n = \begin{cases} \pm\sqrt{v^2 k_x^2 + 2n|b|v} & n \geq 1 \\ \chi sgn(B_y) v k_x & n = 0 \end{cases} \quad (2)$$

in which different chirality $\chi$ correspond to $K$ and $K'$ valleys, respectively. $sgn(B_y)$ denotes the sign of the PMF. Eq. (2) indicates that ZOLMs host linear dispersion but with opposite slopes between $K$ and $K'$ valleys, which can also be evidenced by the numerical calculation in Fig. 2(b). For further clarification, a rectangular array is constructed by periodically repeating the supercell of Fig. 1(a) and its energy spectrum is numerically calculated and exhibited in Fig. 2(c). The ZOLMs (blue dots) are discretely sandwiched between two clusters of higher-order Landau modes (gray regions). Profile of one ZOLM is visualized in Fig. 2(d), in which acoustic wavefunction mainly confines at the intermediate channel of the sample. Its acoustic field distribution along the vertical direction is also quantitatively described by the Gaussian-like distribution as the inset in Fig. 2(d) shows. The ZOLMs exhibit intriguing transported feature distinct from normal bulk states and edge states. This is why we term them as BTs.

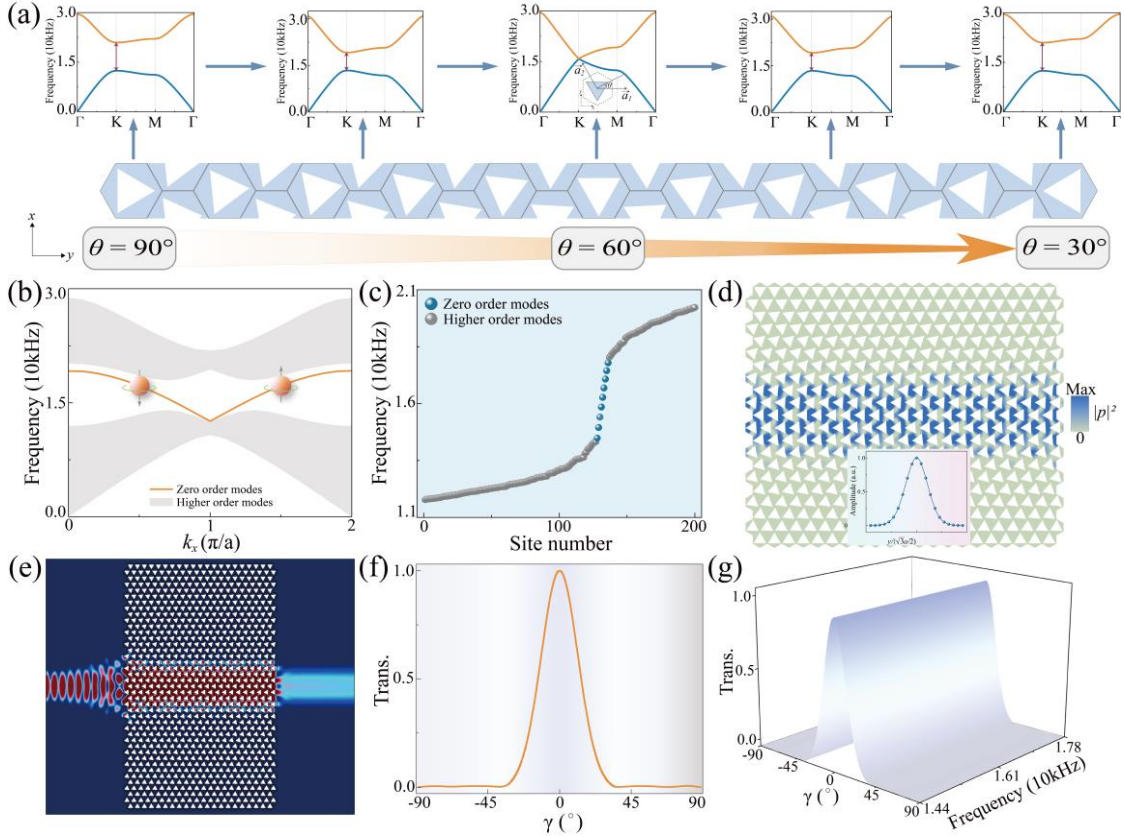

FIG. 2. (a) Design of ribbon-shape supercell with periodicity in $x$ direction while the rotation angle varies gradually from $90^0$ to $30^0$ in $y$ direction. As such, the band structures also evolve and are given at the upper position of corresponding primitive unit. (b) Projected bands of the ribbon-shape supercell, energy spectra (c) and acoustic pressure intensity of zero order modes (d) in 2D acoustic structure for rotation angle $\theta$ covering the range $(30^0, 90^0)$. Similar results are exhibited in (e) - (g) when rotation angle is expanded to cover the range $(30^0, 150^0)$, just twice as wide as the range $(30^0, 90^0)$, but with the same lattice size.

Further, we analyze the transmission properties with the variation of incident angle

$\gamma$ under different frequency $f$ to quantitatively discuss the effect of controlling the BTs in an array with $43 \times 41$ layers. To excite the BTs, a rectangular-shaped acoustic sample is displaced in the air and a Gaussian wave with continuous frequency is imposed at various position of the left edge. Fig. 2(e) presents the transmission spectra as functions of incident angle $\gamma$ and excited frequency $f$. The transmission peak can be achieved at $\gamma = 0$ over the whole spectral gap. It will gradually decrease for increasing incident angle. A slice of the plot at $f = 15000 Hz$ is cut and shown in Fig. 2(f). And the acoustic pressure field for normal incident is exhibited in Fig. 2(g). It clearly indicates that only normally incident wave can excite the BTs and tunnel through the middle bulk channel and exit from the right side. The uniaxial gradient acoustic structure can produce self-collimating effect and filter the obliquely incident wave.

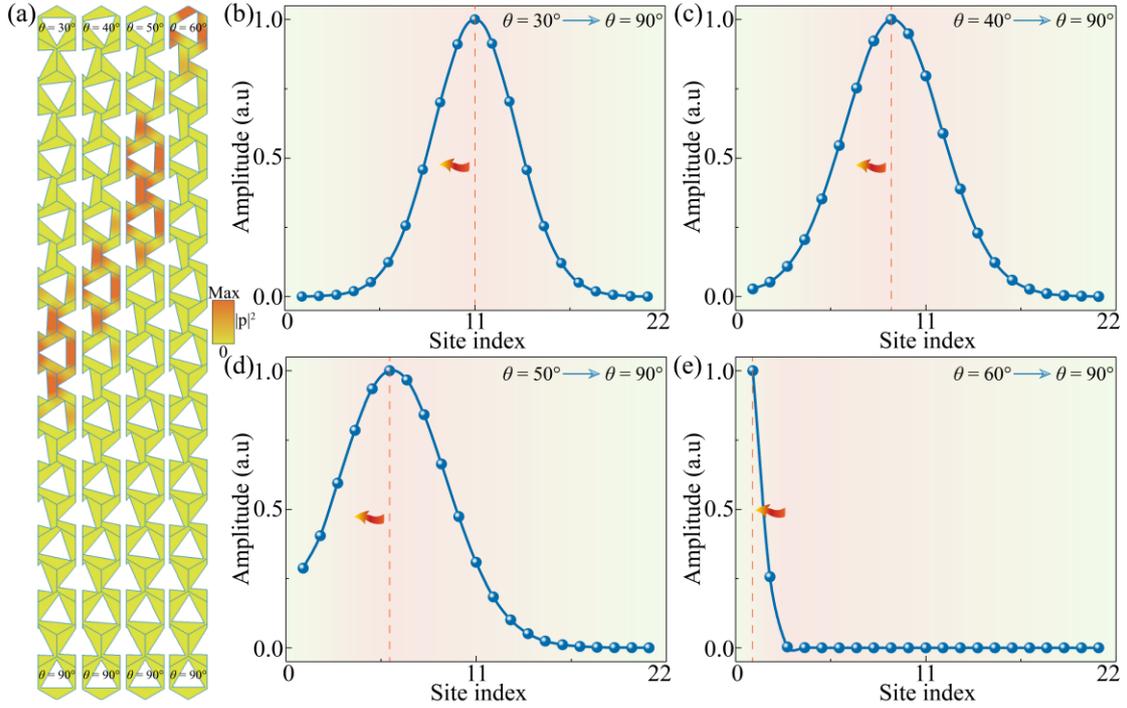

FIG.3. (a) The acoustic pressure distribution in the ribbon-shaped supercells under various rotation angle ranges, corresponding to the cases $(30^0, 90^0)$, $(40^0, 90^0)$, $(50^0, 90^0)$ and $(60^0, 90^0)$. The color denotes the acoustic intensity. (b)-(e) The quantitative plots of acoustic pressure amplitude corresponding to the panels from left to right in (a). The dotted lines sign the maximum position of acoustic field in the supercell. All the quantitatively calculated results are normalized to the maximum value

in the same structure.

From the above study, we know that as BTSs, ZOLMs would bound at the intermediate domain of the gradient lattice when the triangle rotation angle $\theta$ is varied within the range $(30^0, 90^0)$. To better explore its formation mechanism, we further alter the gradient scope of rotation angles. We consider several ribbon-shaped supercells with gradient angles within the ranges $(30^0, 90^0)$, $(40^0, 90^0)$, $(50^0, 90^0)$ and $(60^0, 90^0)$. Periodic boundary conditions are imposed at the left and right edges while the upper and lower ending designated as sound hard boundaries. With the increasing of initial angle, the confined place of the ZOLMs moves toward the ending as the acoustic field exhibit in Fig. 3(a). Bright and dark colors respectively denote the strong and weak acoustic pressure field. The quantitative relation of acoustic pressure and side index is also plotted in Fig. 3(b)-(e) corresponding to the panels from left to right in Fig. 3(a). The numerically calculated results indicate that the bound position of ZOLM mainly dwell at the position of $\theta = 60^0$ with little dependence on the variation range of rotation angle. With the knowledge in mind, we can constitute heterojunction between distinct inhomogeneous lattices for BTSs like the periodic array for conventional edge states.

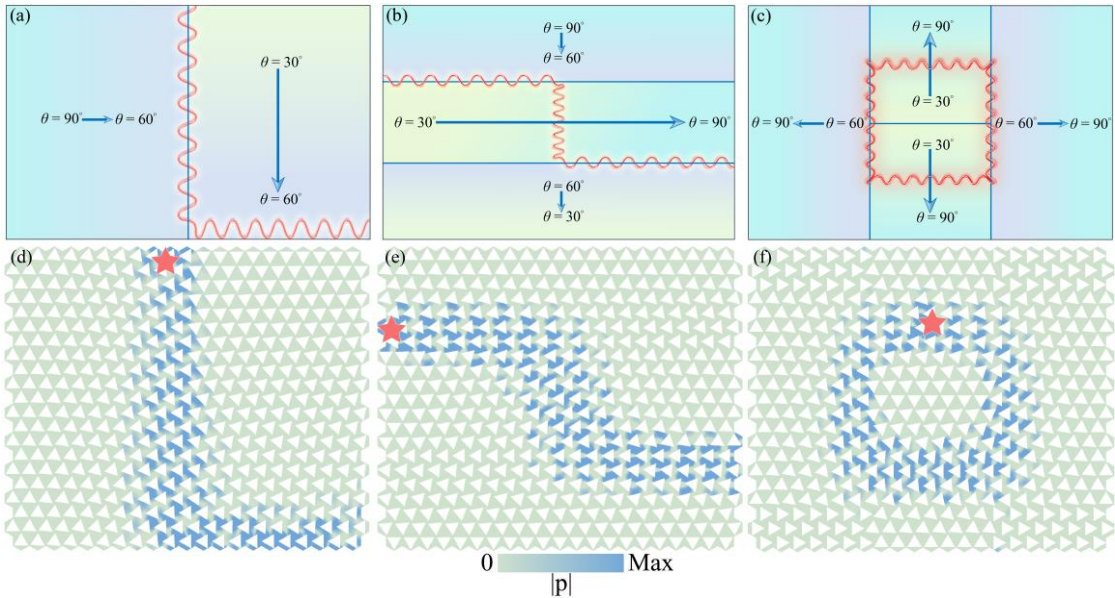

FIG. 4. Construction of different bulk channels for BTSs in an inhomogeneous lattice with configurated pattern. (a)-(c) The schematic diagrams show the realization of "L-

shaped", "Z-shaped" and "O-shaped" routes by tailoring the rotation angle of the blocks. The bulk channels are outlined by wavy lines and separate domains are labelled by arrows to show the gradient direction of rotation angles. (d)-(f) The corresponding acoustic pressure fields with respect to the schemes in (a)-(c) when a point source at frequency $f = 15000\ Hz$ is placed at the position of the channel marked by red stars. The color represents the amplitude of acoustic pressure field.

To validate the new topological route inside the bulk, an inhomogeneous structure is created that covers several domains with different gradient regimes. As the route for BTSs mainly takes place at $\theta = 60^0$, then continuous pathways can be produced by tailoring the variation scope and direction of the gradient angles. We consider several inhomogeneous acoustic domains with gradient angles belonging to the following ranges $(30^0, 90^0), (30^0, 60^0)$ and $(60^0, 90^0)$. By intelligently combining these domains in different patterns, domain walls can be fabricated along the inhomogeneous heterojunction and subsequently bulk channel for BTSs can also be induced. Figs. 4(a)-(c) are the schematic schemes showing the design of bulk channels by engineering the rotation angle patterns. If two gradient domains are assembled in a lattice, L-shaped bulk channel with two passages can be constructed in Fig. 4(a). Similarly, Z-shaped and O-shaped bulk channel can also be achieved by combining three and four aperiodic domains as shown in Figs. 4(b) and 4(c) respectively. When a point source with frequency $f = 15000\ Hz$ is positioned at the bulk channel as the red star mark, BTSs can be excited and transmit along the bulk route. The acoustic pressure field distributions corresponding to the schemes in Figs. 4(a)-(c) are separately displayed in Figs. 4(d)-(f). The realization of BTSs can breakthrough the transport limitation of conventional topological edge states at the edges, making fully use of bulky sample.

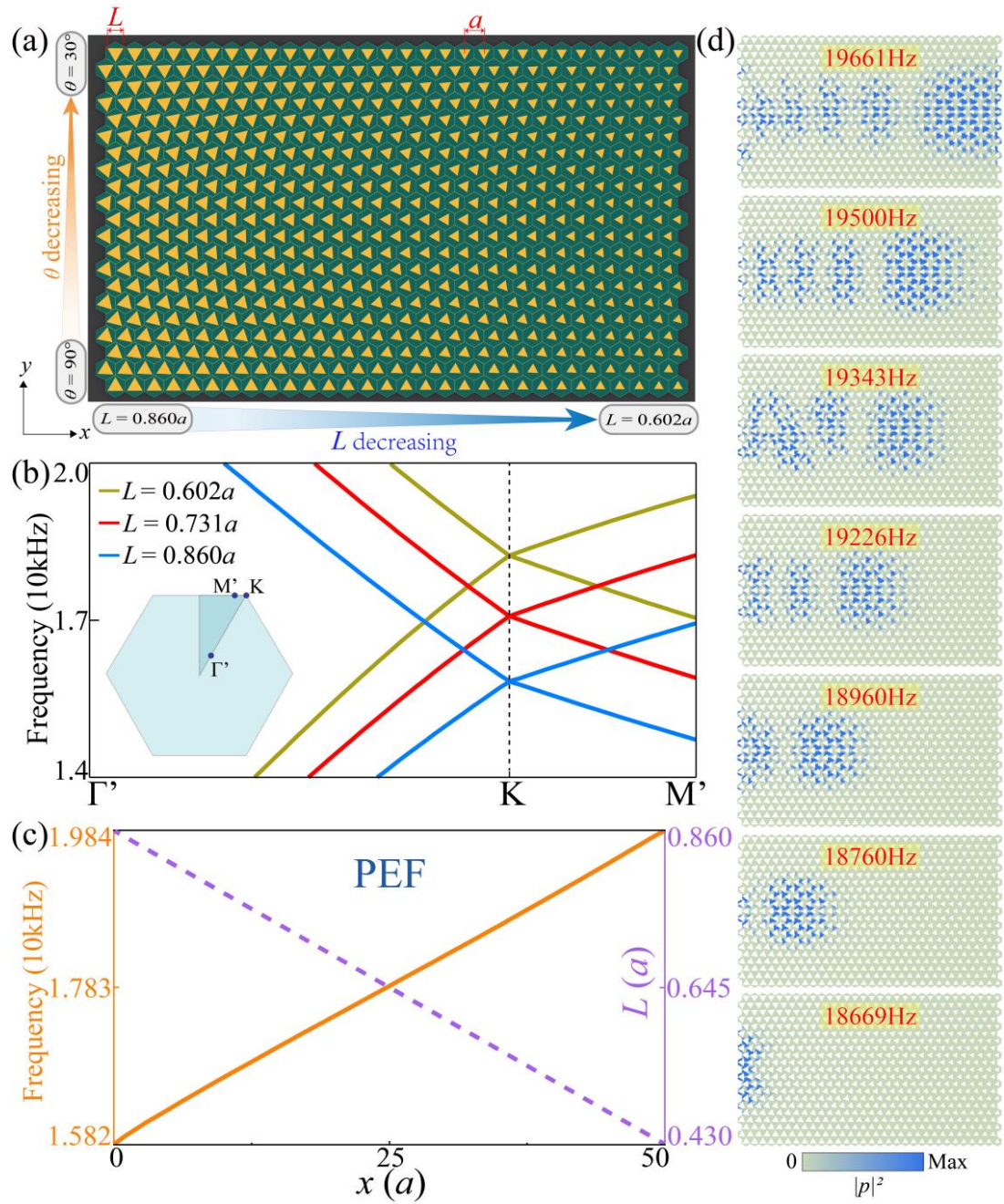

FIG. 5. Realization of PMF and PEF in the same acoustic structure. (a) The 2D sound structure with a hexagonal lattice of triangular scatterers in air background. The configurations of the scatterers are varied in both directions. In $x$ direction, the side length of the triangles $L$ decreases from $0.860a$ to $0.602a$ while rotation angle $\theta$ shifts from $90^0$ to $30^0$ gradually along the $y$ direction. (b) Dispersion relations of the primitive unit cell along the highly symmetric lines $\Gamma' - K - M'$ for side length $L = 0.602a$ (green lines), $0.731a$ (red lines) and $0.860a$ (blue lines). The panel shows the calculated route in the reciprocal space. (c) The evolution of Dirac point and side length of the triangles as a function of distance along $x$ direction. The Dirac frequencies and the side lengths of triangles both exhibit linear dependence on the distance in $x$ direction. (d) The evolution of ZOLM at different frequencies forms Landau rainbow at the central part of 2D aperiodic acoustic structure for $\theta \in (90^0, 30^0)$ and $L \in (0.860a, 0.602a)$.

The above aperiodic structure is axial gradient. Next, we design inhomogeneous acoustic structures with inhomogeneity in two orthogonal directions. As the sketch map in Fig. 5(a) shows, an inhomogeneous acoustic structure is designed with the side lengths and rotation angles of the triangular blocks as two varied physical quantities. In the acoustic array, side lengths of the blocks decrease from left toward right and their rotation angles diminish from bottom to upper. As two independent degrees of freedom, gradually varied side lengths and rotation angles bring different physical consequences. In the above study, we have demonstrated that gradient rotation angles can result in uniform PMFs. Here we will verify that gradient block sizes can induce the other SGF, i.e., uniform PEFs. We first consider three blocks with side length $L = 0.602a, 0.731a$ and $0.860a$ under identical rotation angle $\theta = 0^0$. Their spectral relations along the highly symmetric lines $\Gamma' - K - M'$ in Fig. 5(b) shows that Dirac points move downwards as the increasing of block size. Green, red and blue respectively denote the three blocks from small size to the big one. If block sizes are varied in a sequence, and their side lengths $L$ are linearly dependent on the coordinate $x$. The quantitative relation between side length and coordinate $x$ is $L = -0.0086x + 0.86$ denoted by purple dotted line in Fig. 5(c). Such arrangement of the blocks also produces the Dirac point with linear frequency relation to the coordinate $x$. The concrete mathematic formular is $f = 8040x + 15820$. The frequency change $\delta f$ is equivalent to the scalar potential $\varphi$ and the PEF is given by $\boldsymbol{E} = -\nabla \varphi$. In this case, the uniform PEF amplitude is $E_x = \left|\frac{\delta f}{\delta x}\right| = 8040$. If a PMF $B_y$ and a PEF $E_x$ are included in the same structure like Fig. 5(a) shows, an unprecedented feature called Landau rainbow can emerge inside the bulk channel. In the heterogeneous acoustic sample, there are 50 periods in the $x$ direction and 21 layers in the $y$ direction. The numerically calculated results indicate that ZOLMs move forward along the bulk channel with the ascending of the eigenfrequencies as Fig. 5(d) shows.

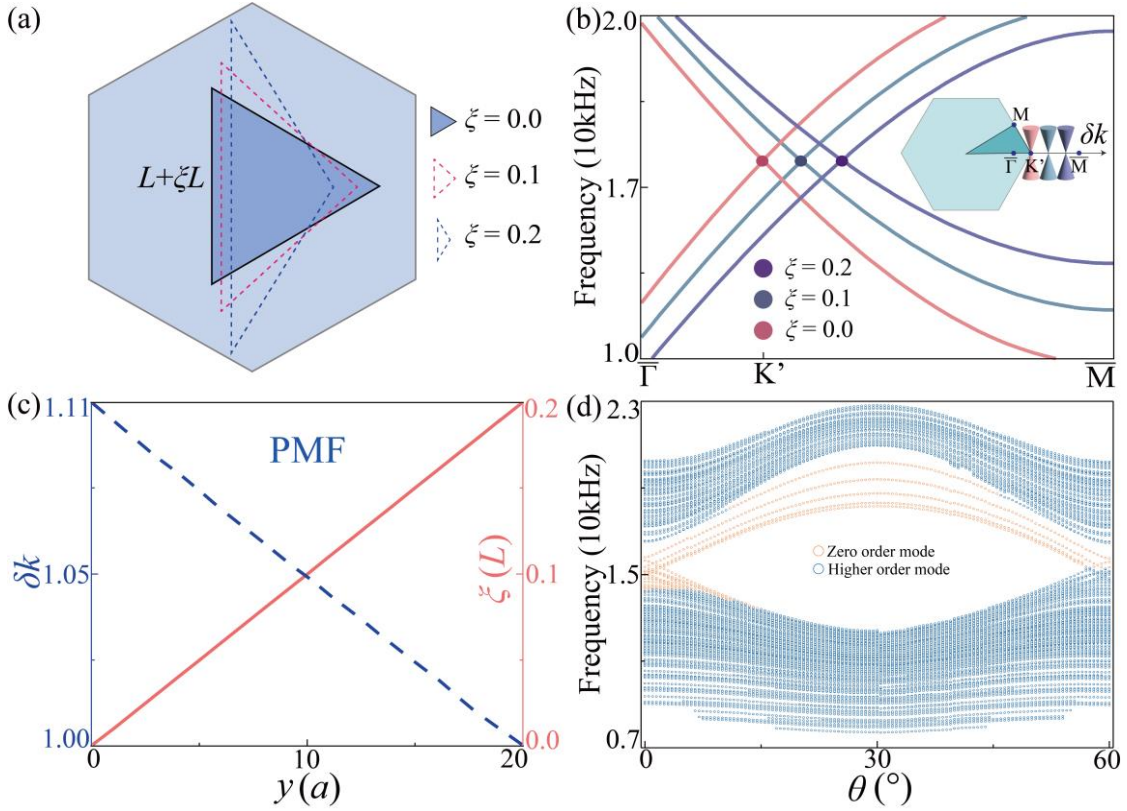

FIG. 6. (a) Schematics of a unit cell with deformed triangles, wherein a side length is characterized by $L(1+\xi)$ while the area of the triangle is constant. (b) By deforming the triangular block, the Dirac point would shift away from original $K$ or $K'$ valley. The inset vividly presents the movement of Dirac cones in the reciprocal space when deformed rate $\xi$ respectively takes the values of $0.0, 0.1$ and $0.2$. (c) The evolution of $\delta k$ (dashed blue line) and deformation rate $\xi$ (red line) in $\Gamma K$ direction as a function of distances along the $y$ direction. (d) The phase diagram shows the Landau modes varied with rotation angle when acoustic blocks are repeated in a ribbon-shaped supercell according to the red line in (c).

In the following, we further consider the third degree of freedom the deformation of the block to construct SGF. The deformation can be attained by stretching or compressing one side of the triangle while its size is fixed. Then the equilateral triangles become isosceles after deformation. The deformed side length is denoted by $L + \xi L$ with $\xi$ denoting the deformation rate. The schematic in Fig. 6(a) shows the block configurations under three deformation cases with $\xi = 0, 0.1$ and $0.2$ denoting by black solid lines, magenta dotted lines and blue dotted lines. Under these deformed rates, their energy bands can be calculated and displayed in Fig. 6(b) when periodic boundary conditions are imposed. Their spectrum shows that as deformation become serious, the Dirac point deviates far more away from initial valley position. The panel shows the first irreducible Brillouin zone and the movement of the Dirac cones along ΓK direction under different deformation. The blocks are arranged in a sequence and their deformation rates have linear relation with respect to the lattice site. The concrete function is $\xi = 8.6y$ as the red curve in 6(c) shows. With this arrangement, the offset of the corresponding Dirac point, $\delta k$, also exhibits a linear relation with the coordinate $y$, which is $\delta k = -5.5y + 1.11$. In the design, a uniform PMF in the z direction is then induced, which is written as $B_z = -A_x/\delta y = \delta k_x/\delta y = 5.5$ according to the relation of magnetic field and magnetic potential $\boldsymbol{B} = \nabla \times \boldsymbol{A}$. If we design a ribbon-shaped supercell with gradually deformed blocks by constant rotation angle along the $y$ direction like the red line in Fig. 6(c) depicts, left and right sides are set as periodic boundary conditions and the remaining edges are sound hard boundaries. Then Landau energy with different Landau orders can be achieved. By varying the rotation angle, the ZOLMs would be remarkably tuned. The phase diagram in Fig. 6(d) shows the relation of the frequencies of zero order (red dots) and other higher order (blue dots) Landau modes and the rotation angle. By examining the profile of Landau modes, we identify that the ZOLMs localize at the boundary of the deformation sample, forming the so-called ETSs. Until now, we recognize that PMF in $xy$ plane and vertical direction $z$ can respectively generate ZOLM in the bulk or at the boundary. This greatly improve the capacity to manipulate the acoustic energy in inhomogeneous background material.

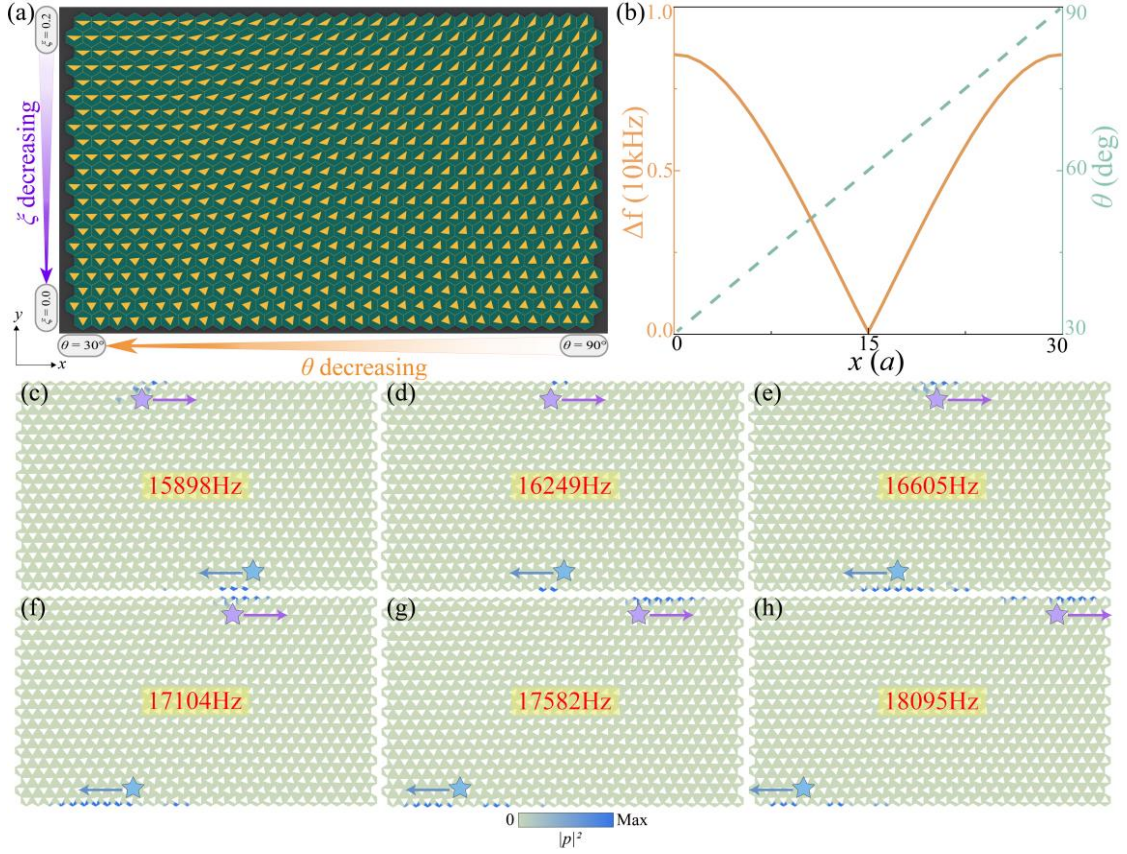

FIG. 7. (a) Construction of uniform PMF and PEF through deforming the triangle configuration and altering rotation angle in the orthogonal directions. (b) Numerically calculated spectrum of the 2D aperiodic acoustic structure in (a). Gray and blue dots respectively indicate higher order and zero order modes induced by Landau level. (c) - (h) The acoustic profiles of zero order modes at different frequencies. The gradually sliding modes at the upper and lower edges form a pair of Landau rainbows, which move in the opposite directions.

In Fig. 5, we have demonstrated that the combination of an in-plane PMF and an in-plane PEF can result in an unprecedented phenomenon termed Landau rainbow, in which BTSs gradually move forward inside the bulk. Here, we verify that the joint action of an out-of-plane PMF and an in-plane PMF can together generate another type of Landau rainbow with ETSs at the boundary. As the acoustic array in Fig. 7(a) shows, the blocks are deformed in the $y$ direction depending on the rule in Fig. 6(c). In the meanwhile, the rotation angles are tuned from $30^0$ to $90^0$ in the $x$ direction as gray dashed line depicts in Fig. 7(b). Under the condition, the complete bandgap $\Delta f$ of higher order Landau modes evolves with position $x$ exactly as red curve shows. When sound hard boundary conditions are imposed on the peripheral edges of the acoustic array, its eigenenergy can be numerically achieved. Within the complete bandgap of the higher order Landau modes, ZOLMs become ETSs. They localize at the upper and lower endings of the array. The upper Landau modes move toward right side and the lower ones move toward left side as the lifting of frequency, forming two counterpropagating Landau rainbows. The acoustic field distribution of ETSs at various frequencies are plotted in Figs. 7(c)-(h). Stars mark the positions of acoustic field and the arrows indicate the moving direction of ETSs. The color represents the intensity of acoustic field as the color bar below indicates.

In conclusion, we have demonstrated that two PMF and a PEF can be successfully implemented by utilizing an acoustic platform. BTSs originate from the in-plane PMF, which is produced by varying rotation angles of the acoustic blocks. And ETSs stem from the in-plane PEF induced by the changed size of the blocks. Both of them are ZOLMs. Furthermore, gradient deformation of the block can also generate out-of-plane PMF. We discover that BTSs have self-collimating effect for incoming waves with various incident angles. If different SGFs are incorporated in the same lattice, an unprecedented phenomenon termed Landau rainbow can be created in the bulk or at the edges of the structure. The phenomenon is the acoustic analog of Landau rainbow of photonic modes realized in [11]. The finding of the SGFs and subsequent Landau rainbow in this work may stimulate the development of acoustic devices with superior performance in energy localization and transport.

## ACKNOWLEDGMENTS

The authors acknowledge support from the National Natural Science Foundation of

China (NNSFC) (Grants No. 12047541, No. 12404493, No. 12274077 and 12404426) and the Research Fund of Guangdong-Hong Kong-Macao Joint Laboratory for Intelligent Micro-Nano Optoelectronic Technology (Grant No. 2020B1212030010).